\documentclass[12pt, a4paper]{article}
\usepackage{amsmath, amssymb}
\usepackage{graphicx}

\usepackage[labelfont=footnotesize,textfont=footnotesize]{caption}   
\usepackage{url}
\usepackage{hyperref}
\usepackage{breakurl}
\usepackage{wrapfig}
\usepackage[square,sort,comma,numbers]{natbib} 
\usepackage{amsmath}
\usepackage{array}
 \usepackage{braket}
\usepackage{mathtools}
\usepackage{titling}
\usepackage{blindtext}
\usepackage{fancyhdr}
\usepackage{appendix}
\usepackage{authblk}
\usepackage[font=small]{caption}

\begin{document}
\hypersetup{colorlinks,%
citecolor=black,%
filecolor=black,%
linkcolor=black,%
urlcolor=blue
}

\date{\normalsize 17 August 2021}
\author{Jordan D. Cohen}
\affil{\it \normalsize Department of Applied Mathematics and Theoretical Physics, Centre for Mathematical Sciences, University of Cambridge, Cambridge, UK}

\title{New Infinities of Soft Charges}
\begin{titlingpage}
    \maketitle
    
    \begin{abstract}
        Recent results on the infrared structure of gravity and electromagnetism have suggested that the deep infrared is much richer than previously appreciated. This article presents a generalisation of these findings within the context of abelian and nonabelian soft (i.e. zero-energy) gauge charges. As a warm up, we describe the emergence of an infinity of soft magnetic charges even in the absence of magnetic monopoles. We show that two infinite sets of soft charges arise in the nonabelian theory as well. In light of the concomitant conservation laws associated with the soft charges, we revisit the black hole information paradox and the no-hair theorems, and argue that a generic black hole carries an infinite amount of gravitational, electromagnetic and chromodynamic soft hair. 
    \end{abstract}
\end{titlingpage}

\numberwithin{equation}{section}

\newpage 
\thispagestyle{empty}

\setcounter{tocdepth}{5}
\newpage
\tableofcontents\protect\enlargethispage{\baselineskip}         

\pagenumbering{arabic}  
\newpage 
\section{Introduction}
The infrared (IR) regimes of gravity and electromagnetism have, in recent years, revealed unexpected relations between three otherwise disparate subjects in physics: soft theorems, asymptotic symmetries, and memory effects. These are summarised by so-called IR triangles, which arise in any physical theory containing gauge degrees of freedom. In brief, an IR triangle characterises an equivalence relation between a class of highly-constrained zero-energy particles, symmetries associated with the asymptotic behaviour of a theory, and a nonlocal memory effect. While each corner of a triangle has provided us with a deeper understanding of its adjoining corners, it is the three-dimensional perspective afforded by a complete triangle that has sourced the most far-reaching of insights. The generators of the asymptotic symmetries, referred to as soft charges, provide perhaps the clearest realisation of this. As they generate continuous asymptotic symmetries, soft charges represent conserved degrees of freedom which reside on the boundaries of a physical system. They are responsible for the production and annihilation of the zero-energy particles described by the soft theorems, and it is the passage of these particles that induces a permanent imprint on certain observables characteristic of a memory effect. The many implications that soft charges and their IR triangles give rise to are still being unravelled today but have thus far included the observations that the vacuum in quantum gravity is infinitely degenerate and that black holes are host to a much lusher head of hair, or information, than previously believed \cite{Strominger:2017zoo}.\footnote{An excellent overview of this which we shall frequently refer to is provided by Strominger's \textit{Lectures on the Infrared Structure of Gravity and Gauge theory} found in \cite{Strominger:2017zoo}.} \\

The first corner of the infrared triangle are the soft theorems. Originating in the context of Quantum Electrodynamics with the works of Bloch and Nordsieck in the early twentieth century, they were eventually generalised to gravity by Weinberg and studied in some detail in the years following \cite{PhysRev.52.54}, \cite{PhysRev.140.B516}. In short, the soft theorems are universal properties of scattering amplitudes and Feynman diagrams that arise whenever an external particle in a scattering process is taken to be \textit{soft}, i.e. low energy. By relating amplitudes with soft particles to amplitudes without, the soft theorems tell us that any physical process can be understood as giving rise to an infinite number of soft particles in a highly regulated way \cite{Strominger:2017zoo}. The second corner of the triangle is the topic of asymptotic symmetries. These are exact symmetries best understood at the asymptotic boundaries of a system. In gauge theories, they are referred to as large gauge transformations. In gravity, we know them as the BMS Symmetries \cite{BMS}. Their deep implications are the subject of ongoing research and include such revelations as the existence of physically significant gauge transformations and the emergence of an infinite-dimensional symmetry group in asymptotically flat spacetimes. The third corner of the triangle concerns the subject of memory. It was first discovered in the context of gravity where it arises as a direct current (DC) effect that describes a permanent change in the positions of inertial bodies due to the passage of gravitational waves. Although we will not discuss the memory effects in this paper, they provide an important physical manifestation of their more abstract IR counterparts and may, in fact, be measurable within the coming decades \cite{Strominger:2017zoo}. \\

Among the most fruitful developments that have emerged from ongoing studies of the infrared is its connection to the black hole information paradox, which arises as an outcome of a semi-classical survey of black hole formation and evaporation. The paradox characterises the loss of information that occurs whenever a black hole forms and subsequently evaporates. The soft charges described by the infrared triangles, however, enforce an infinite number of correlations between the early and late stages of the evaporation process. In other words, black holes carry much more information than arguments for information loss had assumed. While this result is telling, it is still a great deal away from resolving the paradox. Any such solution would require a detailed description of the flow of information between black holes and the external spacetime they inhabit. Doing so would likely reveal novel aspects about the quantum structure of spacetime, and possibly even provide a direct route towards a quantum theory of gravity \cite{Strominger:2017zoo}, \cite{Bojowald:2014zla}, \cite{Hawking:2016msc}. \\

One of the most popular frameworks for studying field theories is the canonical Hamiltonian formalism, within which one constructs a phase space representing the system's possible states along with a set of equations through which dynamical evolutions are expressed. While the formalism is a useful device to examine various features of classical and quantum field theories, it also involves the explicit choice of a time function which renders it non-covariant (no such choice is required for spatial parameters, so space and time are not treated on equal footing in the formalism) \cite{Harlow:2019yfa}. To understand the Hamiltonian dynamics of relativistic field theories one therefore requires a framework that simultaneously preserves covariance while reproducing the phase space of the theory along with its associated structure. The usual approach to this problem, however, is to instead de-emphasise the Hamiltonian formalism and either restrict oneself to Lagrangians in classical theories or path integrals in quantum theories. While both methods have robust explanatory power in their own right, one cannot understate the convenience of a Hamiltonian framework: it is only in this framework that we can properly account for the degrees of freedom of a system – particularly the degrees of freedom that reside on the system's boundaries. This motivated the covariant phase space method due to Iyer, Lee, Zoupas, and Wald, which presents Hamiltonian dynamics in a fully covariant way by preserving the diffeomorphism invariance characteristic of a relativistic field theory \cite{LeeWaldCPSF}, \cite{Wald_2000}.  In this formalism, the symplectic structure of the phase space takes centre stage, and it is by virtue of this structure that we will obtain the principle results of this paper. \\

The text is structured as follows. In Section 2, we present the first complete derivation of the soft magnetic charges of Maxwell's theory. In Section 3, we construct analogous expressions for the soft charges of Yang-Mills theory, which have also lacked a rigorous theoretical treatment to-date, and show that they take an identical form to the soft charges in the abelian theory with the exception of a trace operator that reflects the Lie algebra structure underlying Yang-Mills. The large gauge transformations, or asymptotic symmetries, of both the abelian and nonabelian theories are then described and connected to their accompanying soft charges in Section 4. The black hole information paradox is examined in Section 5 as a central application of the infrared structures discussed in the previous sections. To this extent, we show that the existence of an infinity of soft charges in both classes of gauge theories imposes an infinite number of exact constraints on the formation and evaporation process of a black hole, over and above those already imposed by the soft gravitational charges. We briefly conclude our discussion in Section 6. 

\newpage
\subsection{A lightning overview of the Covariant Phase Space}

To understand the covariant phase space method insofar as it is needed to study the charges that arise in abelian and nonabelian gauge theories, it will be useful to briefly review the subject of symplectic geometry within the context of a field theory. \\

In the Hamiltonian formulation, one studies the dynamics of a phase space labelled by position and momenta coordinates. The phase space is equipped with a smooth scalar function, the Hamiltonian, which generates a dynamical evolution expressed via Hamilton's equations. The separation of the space's coordinates into position and momenta, however, makes it immediately challenging to present the phase space in a way that preserves covariance. The covariant phase space formalism, on the other hand, will take the point of view whereby the phase space is imagined as an abstract manifold equipped with a nondegenerate\footnote{Nondegenerate in this context refers to the fact that the determinant of the two-form expressed as a tensor is nonvanishing or, equivalently, that $\Omega(v, w) = 0 \; \;  \forall \; w \in V$ implies $v = 0$ for an arbitrary vector space $V$ \cite{Seraj:2016cym}.} closed two-form, $\Omega$, known as the symplectic form. The phase space in the covariant formulation is therefore what is referred to as a symplectic manifold \cite{Harlow:2019yfa}, \cite{LeeWaldCPSF}, \cite{Wald_2000}, \cite{AshtekarStreubel}. \\

Although abstract manifolds are, in general, the stage on which most physical processes are studied, it is natural to wonder what utility is provided by a symplectic manifold. Firstly, we wish to construct a Poisson bracket on the phase space from which we can obtain the time evolution of arbitrary functions on the space. One way to define the Poisson bracket of two generic functions on a manifold is via the action of the inverse symplectic form on a variation of the functions, i.e. $\{f, g\} \equiv \Omega^{-1}(\delta f, \delta g)$. The nondegeneracy of the symplectic form, which guarantees the existence of its inverse, is therefore an essential ingredient to obtain well-defined dynamics on the phase space. We then ask that $\Omega$ is closed to ensure that we are able to promote the Poisson bracket of the classical theory into a commutator in the quantum theory \cite{Harlow:2019yfa}.\footnote{There are dynamical systems that do not require that $\Omega$ is closed, but the Poisson bracket is not preserved under the time evolution of such systems so it cannot become a commutator under a canonical quantisation of the theory \cite{Harlow:2019yfa}.} \\ 

\enlargethispage{\baselineskip} \enlargethispage{\baselineskip}  
To arrive at a notion of the symplectic structure of a Lagrangian field theory in a $d$-dimensional spacetime, we note that a variation of the Lagrangian describing a collection of dynamical fields $\phi^{i}$ can always be written as \cite{LeeWaldCPSF} 

\begin{equation}\label{eq: variation of action}
\delta L = \vartheta_{i} \delta \phi^{i} + d\Theta
\end{equation}\

where $\vartheta^{i}$ are the Euler-Lagrange equations, $\delta \phi^{i}$ are the variations of the fields, and $\Theta$ is a $(d-1)$–form known as the presymplectic potential which is linear in both the field variations and its derivatives. To obtain the presymplectic form of the system, $\omega$, we consider a second-order variation of the presymplectic potential \cite{LeeWaldCPSF} 

\begin{equation}\label{eq: generic presymplectic form}
\omega = \delta'\Theta(\delta \phi, \phi) - \delta\Theta(\delta'\phi, \phi)
\end{equation}\

which is manifestly antisymmetric in the variations $\delta$ and $\delta'$. Of course, any product of variations of a 0-form $\phi$ is a two-form which is necessarily antisymmetric in the one-forms, or first-order variations $\delta \phi$, that it is built out of.  While this quantity carries information about the symplectic structure of the phase space we refer to it as presymplectic rather than symplectic since, although it is a closed two-form, it still has degenerate directions that correspond to redundancies in our description of the physical system \cite{LeeWaldCPSF}. We should also point out that the field variations $\delta \phi$ reside in the tangent space of the symplectic manifold so that if $\phi$ a solution to the equations of motion then $\phi + \delta \phi$ is also a solution. In other words, $\delta \phi$ obeys the linearised equations of motion for the field $\phi$.\\

Note that we referred to the presymplectic potential as a $(d-1)$–form yet we regard its second-order variation, the presymplectic form, as a two-form. In fact, the presymplectic potential is a $(d-1)$–form on the spacetime manifold but a one-form on the phase space manifold, while the presymplectic form is a $(d-1)$–form on the spacetime manifold but a two-form on the phase space. The presymplectic potential then becomes a two-form on the phase space after its (second-order) variation is taken as in \eqref{eq: generic presymplectic form}, since the variation operator $\delta$ can be regarded as an exterior derivative on the phase space manifold \cite{Harlow:2019yfa}, \cite{Seraj:2016cym}.\footnote{Note that the requirement that the (pre)-symplectic form is closed is with respect to the phase space exterior differential $\delta$ and not the exterior derivative $d$ on spacetime \cite{Harlow:2019yfa}.}\,\footnote{Hereafter, any object referred to as a $p$-form should be taken as a $p$-form on the spacetime manifold unless otherwise stated.}\\

To obtain the symplectic form of the theory we first impose a gauge fixing on the field variation which eliminates degenerate directions in the presymplectic form. Integrating the gauge-fixed presymplectic form over a (partial) Cauchy surface $\Sigma$ yields the presymplectic potential \cite{LeeWaldCPSF} 

\begin{equation}\label{eq: generic symplectic form}
\Omega = \int_{\Sigma} \omega 
\end{equation}\ \enlargethispage{\baselineskip} \enlargethispage{\baselineskip} \enlargethispage{\baselineskip}\enlargethispage{\baselineskip} \enlargethispage{\baselineskip}

We choose to integrate over a Cauchy surface rather than any arbitrary hypersurface to ensure that our symplectic form captures the Cauchy data of the system \cite{ChoquetBruhat:1969cb}. \\

One of our primary motivations for employing the covariant phase space formalism is that it provides a convenient means to separate the local, or gauge, symmetries of a theory into those that express redundancies in the theory and those that are related to genuine transformations of physical data in the theory. The distinction between these two classes is straightforward: when the generator of a gauge symmetry is trivial then the symmetry is unphysical, i.e. it is a mathematical artifact of our description of the system and no physical observables depend on it \cite{Riello}. If the generator is nontrivial, however, then we are dealing with a so-called large gauge transformation – a physically significant gauge symmetry. We will discuss large gauge transformations in more detail in Section \ref{sec: Asymptotic Symmetries YM} but for now we note that the generators $\mathcal{Q}$ of a generic gauge transformation $\delta A$ obey the relation \cite{BLASCHKE2021115366}

\begin{equation}\label{eq: Gauge Symmetry Gen!}
\delta A  = \{A, \mathcal{Q}\} 
\end{equation}\ 

where $A$ is the connection one-form, or gauge field, charged under the gauge symmetry. The generators of gauge symmetries are what we refer to as first class constraints: constraints on the system's phase space that have a vanishing Poisson bracket with every other constraint on the phase space \cite{Wipf}. Any first class constraint will obey an equation of the form of \eqref{eq: Gauge Symmetry Gen!}. \\

A gauge generator can be explicitly related to the symplectic structure of the theory according to \cite{AshtekarStreubel}

\begin{equation}\label{eq: Gauge generator variation}
\delta \mathcal{Q}  = \int_{\Sigma} \omega
\end{equation}\

In other words, if we construct a symplectic form out of a gauge field $A$ and choose one of the variations of the gauge field to be a gauge transformation, the symplectic form will represent the variation of the gauge generator associated with that transformation. \\

\newpage
\section{Soft Magnetic Charges\label{sec: Soft Magnetic Charges}}
From a technical point-of-view, electrodynamics provides the simplest setting to examine the infrared regime of a quantum field theory which has made the soft electric charges of the theory the subject of considerable interest. Their magnetic counterparts, on the other hand, have received comparatively little attention even while sourcing nontrivial physics in their own right. In this section, we construct an expression for the soft magnetic charge which will serve as a template for deriving the nonabelian soft charges in subsequent sections.\footnote{We work in $(3+1)$-dimensions and adopt natural units throughout this paper, where $\hbar = c = G = 1$. We shall also take all coupling constants to equal 1.} \\

To obtain a formulae for the soft magnetic charges, we begin with an electromagnetic action. The usual approach is to construct such an action as an integral of $F_{\mu \nu}F^{\mu \nu}$ over an arbitrary Lorentzian manifold. The only other possible choices through which one can obtain an electromagnetic action are $(\star F)_{\mu \nu} (\star F)^{\mu \nu}$, which is in fact equivalent to the canonical invariant, and $(\star F)_{\mu \nu} F^{\mu \nu}$ which is associated with magnetic charges \cite{ED}. Although we will not assume that any magnetic sources are present, we will focus on the action corresponding to the latter since it contains the equations of motion associated with magnetic sources. This action is sometimes referred to as topological in the absence of such sources since it is not associated with any dynamical degrees of freedom \cite{YM}. \\

The topological action can be expressed as\footnote{We have used the convention $(\star F)^{\mu \nu} \equiv \epsilon^{\mu \nu \sigma \tau}F_{\sigma \tau}$. Typically, there is a factor of $\frac{1}{2}$ included here but we have omitted this to simplify subsequent formulae.}

\begin{equation}\label{eq: Magnetic Action}
S = - \frac{1}{4} \int_{\mathcal{M}} d^{4}x \sqrt{-g} \, \epsilon_{\mu \nu \alpha \beta} \, F^{\mu \nu} F^{\alpha \beta} 
\end{equation}\

Under a generic variation of the field $A$ $\rightarrow$ $A + \delta A$, it varies as 

\begin{equation}\label{eq: Action Variation by-parts}
\delta S = -\int_{\mathcal{\partial M}} d \Sigma^{\mu} \, \epsilon_{\mu \nu \alpha \beta} \:  \delta A^{\nu} F^{\alpha \beta} + \int_{\mathcal{M}} d^{4}x  \sqrt{-g} \, \epsilon_{\mu \nu \alpha \beta} \: \delta A^{\nu} \nabla^{\mu}F^{\alpha \beta}
\end{equation}\

where the second term vanishes by virtue of the Bianchi identity,

\begin{equation}\label{eq: EOM Magnetic Charges}
\epsilon_{\mu \nu \alpha \beta} \nabla^{\mu}F^{\alpha \beta} = 0
\end{equation}\

The presymplectic potential, a vector field density from which the symplectic structure of the theory can be obtained, is given by the boundary term in \eqref{eq: Action Variation by-parts} 

\begin{equation}\label{eq: Presymplectic Potential for EM}
\theta_{\mu} = - \epsilon_{\mu \nu \alpha \beta} \, \delta A^{\nu} F^{\alpha \beta}
\end{equation}\

According to \eqref{eq: generic presymplectic form} the presymplectic form is

\begin{align}\label{eq: Presymplectic form for EM}
\nonumber \omega_{\mu} & = \delta'\theta_{\mu} (\delta A, F) - \delta \, \theta_{\mu} (\delta ' A, F) \\
& = - \epsilon_{\mu \nu \alpha \beta} (\delta A^{\nu} \nabla^{\alpha} \delta' A^{\beta} - \delta A^{\nu} \nabla^{\beta}\delta' A^{\alpha}  - \delta' A^{\nu} \nabla^{\alpha} \delta A^{\beta} + \delta' A^{\nu}\nabla^{\beta}\delta A^{\alpha})
\end{align}\

Now choosing $\delta' A = \nabla \Lambda$ for the gauge function $\Lambda$, we obtain

\begin{equation}\label{eq: EM Presymplectic form after gauge transf}
\omega_{\mu}  = \epsilon_{\mu \nu \alpha \beta} (\nabla^{\nu}\Lambda \nabla^{\alpha} \delta A^{\beta} - \nabla^{\nu}\Lambda\nabla^{\beta}\delta A^{\alpha})
\end{equation}\

To obtain the symplectic form, and therefore a variation of the charge generating $\nabla \Lambda$, we integrate \eqref{eq: EM Presymplectic form after gauge transf} over a partial Cauchy surface $\Sigma$ to obtain\footnote{There is an important subtlety here: since we have not yet imposed any gauge fixing, \eqref{eq: EM Symplectic form} should still have degenerate directions. This turns out to be an unnecessary detail as long as we restrict ourselves to presymplectic forms that are nonvanishing on $\Sigma$. We will adopt this restriction when deriving the symplectic forms of the nonabelian gauge theory in Section \ref{sec: IR YM} as well.}

\begin{equation}\label{eq: EM Symplectic form}
\delta \mathcal{Q}_{m} = \int_{\Sigma} \omega_{\mu}  d\Sigma^{\mu}  = \int_{\Sigma} \epsilon_{\mu \nu \alpha \beta} (\nabla^{\nu}\Lambda \nabla^{\alpha} \delta A^{\beta} - \nabla^{\nu}\Lambda\nabla^{\beta}\delta A^{\alpha})d\Sigma^{\mu} 
\end{equation}\

Using \eqref{eq: EOM Magnetic Charges} and the fact that the phase space exterior differential $\delta$ commutes with the covariant derivative $\nabla$ (and more formally with the exterior differential on a spacetime manifold, $d$, which follows from their residing in different spaces) we have

\begin{equation}\label{eq: EM Charge variation}
\delta \mathcal{Q}_{m}   = \int_{\Sigma} \epsilon_{\mu \nu \alpha \beta} \nabla^{\nu}(\Lambda (\nabla^{\alpha} \delta A^{\beta} - \nabla^{\beta}\delta A^{\alpha}))d\Sigma^{\mu}
\end{equation}\\

or as a surface integral

\begin{equation}\label{eq: Magnetic Charge variation II}
\delta \mathcal{Q}_{m}    = \int_{\partial \Sigma} \epsilon_{\mu \nu \alpha \beta} \: \Lambda (\nabla^{\alpha} \delta A^{\beta} - \nabla^{\beta}\delta A^{\alpha})d\mathcal{N}^{\mu \nu}
\end{equation}\\

Finally, if the remaining variation of the gauge field $\delta A$ is taken to be pure gauge, we arrive at 

\begin{align}\label{eq: Soft Magnetic Charge}
\nonumber \mathcal{Q}_{m}  &  = \int_{\partial \Sigma}  \epsilon_{\mu \nu \alpha \beta} \: \Lambda F^{\alpha \beta}d\mathcal{N}^{\mu \nu} \\
& = \int_{\partial \Sigma} \Lambda (\star F)^{\mu \nu} \: d\mathcal{N}_{\mu \nu}
\end{align}\

which, in form notation, reads

\begin{equation}\label{eq: Soft Magnetic Charge in forms}
\mathcal{Q}_{m}   = \int_{\partial \Sigma} \Lambda F
\end{equation}\

It immediately follows that any physical theory involving electromagnetism carries an infinite number of ``magnetic charges", one for each choice of $\Lambda$. Since the generator of the gauge transformation is nonvanishing, the gauge transformation is also a genuine physical symmetry.\footnote{More on this in Section \ref{sec: Asymptotic Symmetries YM}.} According to Noether's theorem, the infinity of magnetic charges in \eqref{eq: Soft Magnetic Charge in forms} must therefore obey a corresponding infinity of conservation laws. A common choice is to take $\Lambda$ to be a spherical harmonic on the boundary of the Cauchy surface, i.e. $\Lambda \rvert_{\partial \Sigma} = Y_{ml}$, in which case we would have a distinct magnetic charge for each angular momentum profile $(l, m)$. One can then regard the conservation laws as expressing conservation of magnetic charge at every angle.\footnote{If we chose for $\Sigma$ to be the null-surface boundaries $\mathcal{I}^{\pm}$ so that $\partial \Sigma$ is either $\mathcal{I}^{-}_{+}$ or $\mathcal{I}^{+}_{-}$, then we could interpret \eqref{eq: Soft Magnetic Charge in forms} as the statement that all incoming magnetic quanta on $\mathcal{I}^{-}_{+}$ are equal to all antipodal outgoing magnetic quanta on $\mathcal{I}^{+}_{-}$.} We should be careful to note, however, that unless $\Lambda = 1$, \eqref{eq: Soft Magnetic Charge in forms} are not the magnetic monopoles associated with Dirac quantisation but are rather zero-energy magnetic charges whose existence is independent of the existence of Dirac monopoles. To denominate this distinction, we refer to the magnetic charges in \eqref{eq: Soft Magnetic Charge in forms} as \textit{soft magnetic charges} while referring to Dirac monopoles as \textit{hard magnetic charges}. We shall review in Section \ref{sec: Asymptotic Symmetries YM} a further justification for this naming convention: soft magnetic charges are responsible for the creation and annihilation of soft photons that are associated with the dual of the gauge field.

\newpage

\section{Soft Charges in Yang-Mills}\label{sec: IR YM} 
Generalising the result obtained for electrodynamics to Yang-Mills theory is rather straightforward. We shall work in pure Yang-Mills theory and start by deriving an expression for soft ``electric" charges in Yang-Mills before moving onto the dual soft ``magnetic" charges of the theory.\footnote{Note that this discussion will hold for a general compact, simple Lie Group G.}

\subsection{Soft Chromoelectric Charge}
A simple prescription for deriving the nonabelian counterpart of the soft electric charge is provided by a generalisation of the construction that yielded the soft magnetic charge in the preceding section \cite{Strominger:2017zoo}. We shall refer to it as a soft chromoelectric charge in deference to the fact that the strong force is described by $SU(3)$ Yang-Mills theory. \\

In the absence of matter fields, the dynamical behaviour of a Yang-Mills field is described by the action\footnote{We use the conventions $[T^{a}, T^{b}] = if^{ab}_{\;\; \: c}\,T^{c}$ and $\textrm{tr}(T^{a}T^{b}) = \frac{1}{2}\delta^{ab}$. The gauge covariant derivative is given by $\mathcal{D}_{\mu} = \nabla _{\mu} - i[A_{\mu},\ \ ]$.}\eqref{eq: YM Charge Commutator III} \cite{YM}

\begin{equation}\label{eq: Action YM}
S = -\frac{1}{4}\int d^{4}x \sqrt{-g} \, \textrm{tr}(F_{\mu \nu}F^{\mu \nu}) 
\end{equation}\ 

Under an arbitrary variation of the gauge field, $A \to A + \delta A$, the action varies as

\begin{equation}\label{eq: YM Electro Action Variation II}
\delta S = -\int_{\mathcal{\partial M}} d\Sigma_{\mu} \; \textrm{tr}(\delta A_{\nu} F^{\mu \nu}) + \int_{\mathcal{M}}d^{4}x  \sqrt{-g} \; \textrm{tr}(\mathcal{D}_{\mu}F^{\mu \nu})
\end{equation}\

where the first term contains the presymplectic structure of the theory while the second term vanishes due to the classical equations of motion for the Yang-Mills field \cite{YM}

\begin{equation}\label{eq: EOM YM}
\mathcal{D}_{\mu} F^{\mu \nu} = 0
\end{equation}\ 

We therefore have the following presymplectic potential

\begin{equation}\label{eq: YM Electro Presymplectic potential}
\theta^{\mu} = - \textrm{tr}(\delta A_{\nu} F^{\mu \nu})
\end{equation}\

along with the presymplectic form

\begin{align}\label{eq: YM Electro Presymplectic form}
\nonumber \omega^{\mu} & = - \textrm{tr}(\delta A_{\nu} \delta ' F^{\mu \nu} - \delta' A_{\nu} \delta F^{\mu \nu}) \\ 
& = - \textrm{tr}(\delta A_{\nu} \mathcal{D}^{\mu}\delta'A^{\nu} - \delta A_{\nu}\mathcal{D}^{\nu}\delta'A^{\mu} - \delta'A_{\nu} \mathcal{D}^{\mu}\delta A^{\nu} + \delta'A_{\nu}\mathcal{D}^{\nu}\delta A^{\mu})
\end{align}\

To arrive at a symplectic form for Yang-Mills theory we must integrate \eqref{eq: YM Electro Presymplectic form} over a partial Cauchy surface. 

\begin{equation}\label{eq: YM Electro Symplectic Form}
\Omega  = - \int_{\Sigma} \textrm{tr}(\delta A_{\nu} \mathcal{D}^{\mu}\delta'A^{\nu} - \delta A_{\nu}\mathcal{D}^{\nu}\delta'A^{\mu} - \delta'A_{\nu} \mathcal{D}^{\mu}\delta A^{\nu} + \delta'A_{\nu}\mathcal{D}^{\nu}\delta A^{\mu})d\Sigma_{\mu}
\end{equation}\

As before, after setting one of the variations to a gauge transformation, i.e. $\delta'A_{\mu} = \mathcal{D}_\mu\Lambda$, the symplectic form represents a variation of the charge that generates the gauge transformation. Using the classical equations of motion we have

\begin{equation}\label{eq: YM Chromoelectric charge variation II }
\delta \mathcal{Q}_{\epsilon}   = \int_{\Sigma}\textrm{tr} \:\mathcal{D}_{\nu}\mathcal{G}^{\mu \nu}d\Sigma_{\mu}
\end{equation}\

where $\mathcal{G}^{\mu \nu} \equiv \Lambda (\mathcal{D}^{\mu} \delta A^{\nu} - \mathcal{D}^{\nu}\delta A^{\mu})$. In keeping with the program of Section \ref{sec: Soft Magnetic Charges}, we want to rewrite \eqref{eq: YM Chromoelectric charge variation II } as a surface integral. Naively, one would expect some nonabelian generalisation of Stokes theorem is needed to achieve this, which would involve a notion of ``surface ordering" and reduce \eqref{eq: YM Chromoelectric charge variation II } to an untidy form. Conveniently, however, the action of the gauge covariant derivative on $\mathcal{G}^{\mu \nu}$ reduces to the action of the covariant derivative associated with a spacetime manifold, $\nabla$, since the commutator term that arises from $\mathcal{G}^{\mu \nu}$ taking values in the Lie algebra vanishes. One can check this explicitly by expanding the commutator in the Lie algebra basis

\begin{align}\label{eq: YM Charge Commutator}
\textrm{tr}([A_{\nu},\mathcal{G}^{\mu \nu}]) & = \textrm{tr}(\Lambda [A_{\nu}, \delta F^{\nu \mu}] + [A_{\nu}, \Lambda]\delta F^{\nu \mu}] \\
& = \textrm{tr}(\Lambda_{d} A_{\nu, a}\delta F^{\nu \mu}_{b} T^{d}[T^{a},T^{b}] + \Lambda_{b}A_{\nu, a} \delta F^{\nu \mu}_{d}[T^{a}, T^{b}]T^{d})
\end{align}\

Upon relabelling indices in the second term we are left with 

\begin{equation}\label{eq: YM Charge Commutator III}
\textrm{tr}([A_{\nu},\mathcal{G}^{\mu \nu}]) = \frac{i}{2}(\Lambda_{c} A_{\nu, a}\delta F^{\nu \mu}_{b} f^{abc} + \Lambda_{c} A_{\nu, a}\delta F^{\nu \mu}_{b} f^{acb})
\end{equation}\

which vanishes due to the anti-symmetry of the fine structure constant $f^{acb} = -f^{abc}$. \\

The variation of the charge $\mathcal{Q}$ is therefore

\begin{equation}\label{eq: YM Chromoelectric charge variation III }
\delta \mathcal{Q}_{\epsilon}   = \int_{\Sigma}\textrm{tr} \nabla_{\nu}\mathcal{G}^{\mu \nu} d\Sigma_{\mu}
\end{equation}\

which degenerates into the surface integral

\begin{equation}\label{eq: YM Chromoelectric charge variation surface }
\delta \mathcal{Q}_{\epsilon}   = \int_{\partial \Sigma}\textrm{tr}\, \mathcal{G}^{\mu \nu} d\mathcal{N}_{\mu \nu}
\end{equation}\

If the variation  $\delta F^{\mu \nu}$ implicit in $\mathcal{G}^{\mu \nu}$ is the difference between the null-field configuration and some specific field configuration $F^{\mu \nu}$ then we obtain an expression for the charge that is linear in the field strength

\begin{equation}\label{eq: Soft chromoelectric charge}
\mathcal{Q}_{\epsilon}  = \int_{\partial \Sigma}\textrm{tr}\Lambda F^{\mu \nu}\, d\mathcal{N}_{\mu \nu} 
\end{equation}\

or, in form notation,

\begin{equation}\label{eq: Soft chromoelectric charge II}
\mathcal{Q}_{\epsilon}   = \int_{\partial \Sigma}\textrm{tr} \, \Lambda \star F
\end{equation}\

With the exception of the trace operation, the expression for soft chromoelectric charge is identical in form to the expression for soft electric charge. We have therefore found that, in Yang-Mills theory for a compact Lie Group $G$, an infinite set of soft chromoelectric charges emerge. Since the charges are generators of continuous symmetries, they each obey a conservation law which tightly constrains any physical process involving Yang-Mills fields. Under an appropriate choice for the set of $\Lambda$ on $\partial \Sigma$, such as the set of spherical harmonics, the conservation equations can be regarded as expressing the conservation of chromoelectric charge at every angle. 

\subsection{Soft Chromomagnetic Charge}
To obtain an analogous expression for the dual soft charges of the theory, or soft chromomagnetic charges, we compute a variation of the topological action of Yang-Mills. \\

The action in question is given by\footnote{We again assume the absence of matter fields.}

\begin{equation}\label{eq: Topological Action YM}
S = -\frac{1}{4}\int d^{4}x \sqrt{-g} \, \textrm{tr}(\epsilon_{\alpha \beta \mu \nu} \, F^{\alpha \beta}F^{\mu \nu}  ) 
\end{equation}\ 

Unlike \eqref{eq: Action YM}, we do not refer to this action as describing the dynamical behaviour of Yang-Mills fields since it is not associated with any dynamical equations of motion for the fields. \\

Under $A_{\mu} \to A_{\mu} + \delta A_{\mu}$ the variation of the topological action can be written as

\begin{equation}\label{eq: YM Topo Action Variation Surface II}
\delta S = -\int_{\mathcal{\partial M}} \textrm{tr}(\epsilon_{\alpha \beta \mu \nu}\,\delta A^{\beta}F^{\mu \nu})d\Sigma^{\alpha} + \int_{\mathcal{M}} d^{4}x \sqrt{-g} \, \textrm{tr}(\delta A_{\rho}\epsilon_{\alpha \beta \mu \nu}\, g^{\rho \beta} \mathcal{D}^{\alpha} F^{\mu \nu} )
\end{equation}\

where the bulk term vanishes since the inner bracketed term is the Bianchi identity

\begin{equation}\label{eq: Bianchi II}
\mathcal{D}_{\mu}(\star F)^{\mu \rho} = \epsilon_{\alpha \beta \mu \nu}\, g^{\rho \beta}\mathcal{D}^{\alpha}F^{\mu \nu} = 0
\end{equation}\

As before, the presymplectic potential is contained within the boundary term in \eqref{eq: YM Topo Action Variation Surface II} 

\begin{equation}\label{eq: Mag YM Presymplectic Potential}
\theta_{\alpha} = -\textrm{tr}(\epsilon_{\alpha \beta \mu \nu}\,\delta A^{\beta}F^{\mu \nu})
\end{equation}\

The corresponding presymplectic structure form is

\begin{align}\label{eq: Presymplectic Form YM Mag}
\nonumber \omega_{\alpha} & = -\textrm{tr}(\epsilon_{\alpha \beta \mu \nu}\,(\delta A^{\beta}\delta'F^{\mu \nu} - \delta' A^{\beta}\delta F^{\mu \nu})) \\
&  = - \textrm{tr}(\epsilon_{\alpha \beta \mu \nu}\,(\delta A^{\beta}(\mathcal{D}^{\mu}\delta'A^{\nu} - \mathcal{D}^{\nu}\delta' A^{\mu}) - \delta' A^{\beta}(\mathcal{D}^{\mu}\delta A^{\nu} - \mathcal{D}^{\nu}\delta A^{\mu}))
\end{align}\

If $\delta ' A^{\nu} = \mathcal{D}^{\nu}\Lambda$ then the bracketed term on the left vanishes and, upon integrating over an arbitrary manifold $\Sigma$, we are left with a variation of the charge that generates this gauge transformation

\begin{equation}\label{eq: Charge variation YM Mag}
\delta \mathcal{Q}_{m} = \int_{\Sigma} \textrm{tr}(\epsilon_{\alpha \beta \mu \nu}\mathcal{D}^{\beta}\Lambda \, \delta F^{\mu \nu})d\Sigma^{\alpha}
\end{equation}\

where $\delta F^{\mu \nu} = \mathcal{D}^{\mu}\delta A^{\nu} - \mathcal{D}^{\nu}\delta A^{\mu}$. Just as in the case of the soft magnetic charge, one can turn \eqref{eq: Charge variation YM Mag} into a total derivative since the exterior covariant derivative $\mathcal{D}$ acting on the variation of the field strength in \eqref{eq: Charge variation YM Mag} is just the variation $\delta$ on the Bianchi identity.\footnote{To be precise, it is the variation of the Bianchi identity up to a sign.} We are therefore left with 

\begin{equation}\label{eq: Charge Variation YM Mag II}
\delta \mathcal{Q}_{m} = \int_{\Sigma} \textrm{tr}(\epsilon_{\alpha \beta \mu \nu}\, \mathcal{D}^{\beta}\mathcal{G}^{\mu \nu})d\Sigma^{\alpha}
\end{equation}\

which we have simplified with $\mathcal{G}^{\mu \nu} \equiv \Lambda\delta F^{\mu \nu}$. Since the commutator contained within the gauge covariant derivative on $\mathcal{G}^{\mu \nu}$ vanishes as it did in \eqref{eq: YM Charge Commutator III}, we can apply the usual form of Stokes theorem to obtain

\begin{equation}\label{eq: Charge Variation YM Mag III}
\delta \mathcal{Q}_{m} = \int_{\partial \Sigma} \textrm{tr}(\epsilon_{\alpha \beta \mu \nu}\, \mathcal{G}^{\mu \nu})d\Sigma^{\alpha \beta}
\end{equation}\

Finally, if the variation of the field strength $\delta F^{\mu \nu}$ is the difference between the null field configuration where the field strength is vanishing and an arbitrary nonvanishing field configuration, then we arrive at

\begin{equation}\label{eq: Soft Chromomagnetic Charge}
\mathcal{Q}_{m} = \int_{\partial \Sigma} \textrm{tr} \,\epsilon_{\alpha \beta \mu \nu} \, \Lambda F^{\mu \nu} \; d\Sigma^{\alpha \beta}
\end{equation}\

Which in form notation reads

\begin{equation}\label{eq: Soft Chromomagnetic Charge form}
\mathcal{Q}_{m} = \int_{\partial \Sigma} \textrm{tr} \, \Lambda F
\end{equation}\

Yang-Mills theory therefore contains an infinity of soft chromomagnetic charges. As before, there is a concomitant infinity of conservation equations which have the same physical interpretation as the conservation equations of the nonabelian soft charges and which restrain any physical process in which the Yang-Mills fields play a role. We shall see in Section \ref{sec: Soft Implants} that the existence and conservation of these quantities has important implications for black hole physics. 
\newpage

\section{Asymptotic Symmetries in Gauge Theories\label{sec: Asymptotic Symmetries YM}}

We have briefly mentioned that soft charges are, in general, related to physically significant gauge transformations of the gauge fields – particularly of the field configurations that reside on the boundaries of the system. To appreciate this statement, however, we first need to clarify what we mean by the boundaries of a physical system. \\

In the covariant phase space formalism, for any field theory in a spacetime that asymptotically approaches Minkowski spacetime one ought to choose a surface that all Cauchy surfaces $\Sigma$ have as their boundary\footnote{We restrict ourselves to the case that such a boundary is connected.} along with appropriate boundary conditions for the fields \cite{Kirklin:2018wvq}. Since the spacetime we consider is infinite, we imagine a conformal compactification of the spacetime which affords us with a natural means to study the structure of its boundaries. Under such a compactification, the boundary is comprised of five components:

\begin{itemize}
\item Future/past timelike infinity, $i^{\pm}$, which corresponds to the infinite future/past of any massive fields in the spacetime. 
\item Spacelike infinity, $i^{0}$, which is an infinite spacelike distance from any point in the spacetime. Spacelike infinity is causally disconnected from the rest of the spacetime and therefore cannot accommodate any observers. 
\item Future/past null infinity, $\mathcal{I}^{\pm}$, which are the future/past endpoints of any null curves in the spacetime. \\
\end{itemize}

While the most common choice for a boundary $\partial \Sigma$ is $i^{0}$, we are interested in the nature of the field configurations at $\mathcal{I}^{\pm}$ since neither radiative modes nor matter fields exist at $i^{0}$. Indeed, it is at $\mathcal{I}^{\pm}$ where massless excitations end up and therefore where IR triangles arise. \\

We begin our analysis of the connection between soft charges and gauge transformations at $\mathcal{I}^{\pm}$ by imposing simplifying restrictions that will enable us to study the essential features of this connection. First, we restrict our focus to the structure on $\mathcal{I}^{+}$ and ignore the counterpart structure on $\mathcal{I}^{-}$. In this spirit, we assume that there are no incoming hard charges, i.e. any current or charge densities vanish on $\mathcal{I}^{-}$. \\

It is convenient to use retarded Bondi coordinates to parameterise $\mathcal{I}^{+}$. In doing so, we parameterise the spatial component of the spacetime with complexified spherical coordinates $(r, z, \bar{z})$ such that the line element in the asymptotic limit of the spacetime is given by \cite{Strominger:2013lka}

\begin{equation}\label{eq: Spacetime metric}
ds^{2} = -du^{2} - 2dudr + 2r^{2}\gamma_{z\bar{z}}dzd\bar{z}
\end{equation}\
 
with $u = t - r$ and $z = e^{i\phi}\,  \textrm{tan}{\tfrac{\theta}{2}}$. The metric on the unit sphere is then $\gamma_{z\bar{z}} \, = 2/(1+ z \bar{z})^{2}$ \cite{Strominger:2013lka}.\\
 
Now recall that actions of both abelian and nonabelian gauge fields carry a gauge symmetry given by \cite{ED}

\begin{equation}\label{eq: Gauge Symmetry ED}
\delta A_{\mu} = \mathcal{D}_{\mu}\Lambda
\end{equation}\

where $\mathcal{D}_{\mu}$ is a gauge covariant derivative in the nonabelian case and a partial derivative in the abelian case. \\

It is convenient to study the system under the gauge fixing condition

\begin{equation}\label{eq: Temporal Gauge ED}
A_{u} = 0
\end{equation}\

which is often referred to as the temporal gauge condition \cite{Strominger:2017zoo}. This is only a partial gauge fixing, however, since there is still a residual $u$-independent gauge symmetry that remains. That is to say, 

\begin{equation}\label{eq: Residual Gauge Redundancy ED}
A_{u}' = A_{u} + \partial_{u}\Lambda 
\end{equation}\

still satisfies \eqref{eq: Temporal Gauge ED} provided that the gauge function is independent of retarded time. We impose an additional partial gauge fixing by demanding that, at $\mathcal{I}^{+}$, we have 

\begin{equation}\label{eq: Radiation Gauge ED}
r^{2}\gamma_{z \bar{z}} \nabla^{\mu}A_{\mu}\bigg\rvert_{\mathcal{I}^{+}} = \partial_{z}A_{\bar{z}} + \partial_{\bar{z}}A_{z} + (\partial_{r} - \partial_{u})(\gamma_{z \bar{z}}r^{2}A_{r})\bigg\rvert_{\mathcal{I}^{+}} = 0 
\end{equation}\

The above equation is usually referred to as the radiation gauge \cite{Strominger:2013lka}. The gauge transformations that still remain are

\begin{align}\label{eq: Residual Gauge Transformations ED}
  \nonumber \partial_{u}\Lambda = 0 \\
 2 \partial_{z}\partial_{\bar{z}}\Lambda + \gamma_{z \bar{z}} \partial_{r}(r^{2}\partial_{r}\Lambda) = 0 
\end{align}\

The solution to \eqref{eq: Residual Gauge Transformations ED} is given by the holomorphic function 

\begin{equation}\label{eq: Residual gauge function ED}
\Lambda = \lambda(z) + \bar{\lambda}(\bar{z})
\end{equation}\ 

Locally, this leaves the residual transformation

\begin{equation}\label{eq: Residual transformation ED}
\delta A_{z} = \mathcal{D}_{z}\lambda
\end{equation}\

which tells us that the transformation has a nontrivial action on the radiative mode $A_{z}$ of the gauge field.  Of course, globally there is no holomorphic function on the 2-sphere at null infinity (and on any 2-sphere for that matter) except for the case that $\Lambda = const.$, which implies that \eqref{eq: Residual transformation ED} contains singularities at the poles \cite{Strominger:2013lka}. \\

We are now able to appreciate the physical significance of the residual transformation in \eqref{eq: Residual transformation ED}. Naively, one would expect that all gauge transformations are unphysical. In the case that one has a Cauchy surface with nontrivial boundary, however, the charge generating gauge symmetries that are implemented at that boundary is nonvanishing \cite{Kirklin:2018wvq}. In light of \eqref{eq: Gauge Symmetry Gen!} and \eqref{eq: Residual transformation ED} we therefore have, at $\mathcal{I}^{+}$, that \cite{Hawking:2016msc} 

\begin{equation}\label{eq: Effect of Soft Electric Charges}
[\mathcal{Q}_{\epsilon}, A_{z}(u, z, \bar{z})] = i \mathcal{D}_{z}\lambda
\end{equation}\

where $\mathcal{Q}_{\epsilon}$ refers to the soft (chromo)-electric charges of the relevant theory.\footnote{We will motivate in a moment why the soft charge here is ``electric" and not ``magnetic".} One may interpret \eqref{eq: Effect of Soft Electric Charges} as expressing the fact that the radiative mode of the gauge field at the boundary of the spacetime transforms nontrivially under the action of the soft charge generating the gauge symmetry. In fact, \eqref{eq: Effect of Soft Electric Charges} is an expression of spontaneous symmetry breaking: the residual gauge symmetry \eqref{eq: Residual transformation ED} is spontaneously broken due to the presence of a nontrivial boundary. The corresponding ``boundary" Goldstone bosons are the zero modes of $A_{z}$ whose polarisation vectors are proportional to $\mathcal{D}_{z}\lambda$ \cite{Strominger:2017zoo}. Indeed, after promoting $\mathcal{Q}_{\epsilon}$ to an operator, we have \cite{Hawking:2016msc}

\begin{equation}\label{eq: Soft Charges as Operators}
\mathcal{Q}_{\epsilon}|0 \rangle = \left( \int_{\partial \Sigma} \Lambda \star F \right)|0 \rangle \neq  0 
\end{equation}\

for a vacuum state $|0 \rangle$. In other words, the action of the theory is preserved under the symmetries generated by $\mathcal{Q}_{\epsilon}$ but the vacuum structure is not: there is a nonvanishing quantity associated with the vacuum which is labelled by the infinite set of quantum numbers accompanying \eqref{eq: Soft Charges as Operators}. We are therefore able to move between physically-distinct vacua by acting on the vacuum with soft charges or, equivalently, by creating Goldstone bosons on the spacetime boundary. In the abelian case the Goldstone bosons are zero-energy photons while in the nonabelian case they are zero-energy gluons.\\ 

Any gauge transformations which are spontaneously broken due to the presence of a nontrivial boundary are referred to as large gauge transformations; they are the gauge analogues of the BMS supertranslation symmetries of gravitational theories. Gauge symmetries which are unphysical, on the other hand, are referred to as small gauge transformations \cite{Kirklin:2018wvq}. \\

To understand why \eqref{eq: Effect of Soft Electric Charges} characterises only ``electric" large gauge transformations we will need to consider the situation from the perspective afforded by Maxwell's theory.\footnote{A general gauge theoretic point-of-view would not be as useful for this analysis since a well-defined derivation of ‘’magnetic” large gauge transformations in the nonabelian theory is still lacking. This is since, in Yang-Mills theory, it does not appear to be the case that a dual gauge field can be straightforwardly constructed as it can in the abelian theory. It seems reasonable to assume, however, that a nonabelian generalisation of the magnetic large gauge transformations of Maxwell's theory is still possible. We leave this to future investigation.} In this case, our analysis concerned the asymptotic symmetries of the gauge field $A$ belonging to the field strength $F$ from which an action associated with electric sources can be obtained. We could have instead performed the analysis with the dual gauge field $\tilde{A}$ from which the dual field strength $\star F$ is built. Rather than repeating the above procedure, however, we will take advantage of a simple trick. First, we define the dual variable \cite{Strominger:2017zoo}

\begin{equation}\label{eq: Dual Variables}
\tilde{F} = - \star F
\end{equation}\

which is related to the dual gauge field via

\begin{equation}\label{eq: Dual Field with Dual Gauge}
\tilde{F} = d\tilde{A} = - \star dA
\end{equation}\

In the presence of hard magnetic charges, the dual gauge field couples to magnetic sources in the same way that the gauge field couples to electric sources. As with the gauge field, \eqref{eq: Dual Field with Dual Gauge} is defined up to a (dual) gauge transformation \cite{Strominger:2017zoo}. \\

In order to obtain a magnetic analogue of \eqref{eq: Residual transformation ED}, we must examine \eqref{eq: Dual Field with Dual Gauge} near $\mathcal{I}^{+}$. To a first-order approximation, we have \cite{Strominger:2017zoo}

\begin{equation}\label{eq: Dual field on null infinity}
\tilde{F}_{uz}\, \Bigg\rvert_{\mathcal{I}^{+}} = i \, F_{uz}\, \Bigg\rvert_{\mathcal{I}^{+}}
\end{equation}\ 

After integrating \eqref{eq: Dual field on null infinity} we find \cite{Strominger:2017zoo}

\begin{equation}\label{eq: Dual Gauge Field relation}
\tilde{A}_{z} =  i A_{z}
\end{equation}\ 

This is a suggestive equation – it says that the dual gauge field is simply the gauge field shifted by the phase $\frac{\pi}{2}$. The analogue of \eqref{eq: Residual transformation ED} for the dual field follows immediately, yielding 

\begin{equation}\label{eq: Residual transformation Mag}
\delta \tilde{A}_{z} =  i \partial_{z}\lambda
\end{equation}\

which implies that 

\begin{equation}\label{eq: Effect of Soft Magnetic Charges}
[\mathcal{Q}_{m}, \tilde{A}_{z}(u, z, \bar{z})] = \partial_{z}\lambda
\end{equation}\

We may therefore interpret the effect of the soft magnetic charge $\mathcal{Q}_{m}$ generating the above residual gauge transformation in exactly the same way as we did for the soft electric charge: the radiative mode of the dual, or phase-shifted, gauge field transforms nontrivially at the spacetime boundary under the action of $\mathcal{Q}_{m}$. The Goldstone bosons of the spontaneously broken gauge symmetry described by \eqref{eq: Residual transformation Mag} are hence soft photons that differ from their electric counterparts by the phase $\frac{\pi}{2}$. In other words, $\mathcal{Q}_{m}$ results in soft photons whose electric and magnetic fields are exchanged relative to the soft photons associated with $\mathcal{Q}_{\epsilon}$. This is a statement about electromagnetic helicity. To see this we note that, in the absence of sources, the electromagnetic field carries the so-called duality symmetry \cite{Calkin:1965} 

\begin{align}\label{eq: Duality transf}
& \nonumber \mathbf{E} \to \mathbf{E'} = \mathbf{E}\, \textrm{cos}\theta + \mathbf{B} \, \textrm{sin}\theta \\
& \mathbf{B} \to \mathbf{B'} = -\mathbf{E}\, \textrm{sin}\theta + \mathbf{B} \, \textrm{cos}\theta 
\end{align}\

which rotates the electric field $\mathbf{E}$ into the magnetic field $\mathbf{B}$ and vice versa. The conserved quantity associated to this transformation, which was first recognised by Calkin in 1965, is the electromagnetic helicity \cite{Calkin:1965}. Since the relative exchange of the electric and magnetic fields between $\mathcal{Q}_{m}$ and $\mathcal{Q}_{\epsilon}$ is just the duality rotation in \eqref{eq: Duality transf} with $\theta = \frac{\pi}{2}$, the soft photons due to $\mathcal{Q}_{m}$ necessarily carry the same helicity as the soft photons due to $\mathcal{Q}_{\epsilon}$. \\

In summary, the soft charges of Maxwell's theory are the generators of physically nontrivial gauge transformations that create and annihilate soft photons on the spacetime boundary. The only distinction between the soft photons generated by the electric and magnetic species is a relative exchanging of the electric and magnetic fields. In the nonabelian scenario, \eqref{eq: Effect of Soft Electric Charges} reflects the fact that soft chromoelectric charges generate gauge transformations that create and annihilate soft modes of the nonabelian gauge field, or soft gluons, on the spacetime boundary. A similar result for soft chromomagnetic charges is expected, although novel differences to the abelian case may exist.\footnote{For one, Yang-Mills theory does not carry a duality symmetry so zero-energy gluons created by soft chromomagnetic charges do not necessarily carry the same helicity as those created by soft chromoelectric charges.}
\newpage

\section{Soft Implants on Black Holes\label{sec: Soft Implants}}
Research on infrared physics has, in recent years, resulted in a flurry of works that have revealed a structure rich with insights into the nature of gravity, not least through the study of black hole physics. In spite of this, the information paradox remains a major obstacle in our development of a full quantum theory of gravity. It is here that the soft gauge charges, along with their gravitational analogues first presented in \cite{Hawking:2016msc} and \cite{Hawking:2016sgy}, find perhaps their most salient physical application.\\

To appreciate the relevance of soft charges in the information paradox, one should first attempt to understand precisely what the paradox is. This is easier said than done, not because the information paradox is incomprehensible but because it makes itself manifest in such a variety of ways. It is sufficient for the purposes of this analysis to characterise one such way: the formation and (presumably) total evaporation of a black hole in an asymptotically Minkowskian spacetime.\footnote{We will justify this presumption shortly.}\\

From the classical perspective, any generic black hole will eventually settle down to a state characterised by just ten conserved charges, commonly known as the Poincaré charges. These are the energy-momentum of the system, its total angular momentum, and the three generators associated with spacetime boosts \cite{Robinson:2004zz}. A black hole in Einstein gravity therefore has ten hairs, i.e. its phase space is ten-dimensional \cite{Strominger:2017zoo}.\footnote{In the presence of gauge fields there will be additional conservation laws that add hairs to a black hole, such as conservation of total electric charge, but this is not important for this discussion.} This is the so-called no hair theorem.\\

Upon surveying a black hole from a quantum mechanical point-of-view, the situation becomes rapidly more nuanced. Indeed, after a finite period of time and assuming that no Planck scale remnants occur, particle production near the event horizon will necessarily result in the black hole's total evaporation \cite{Hawking:1974sw}.\footnote{Although Planck scale remnants have been proposed as a possible solution to the information paradox they are not without their own problems. Whether or not they interact with the outside world presents a particularly crucial example. If there is no interaction, then they don't appear to be any good at solving the paradox since they would not preserve unitarity and therefore information. If they do interact with the outside world then they become entropically-favoured over all other forms of matter which implies that they should occur spontaneously and in considerable numbers \cite{Unruh:2017uaw}.} The key point is that if the collapsing matter began in a pure quantum state then the state after total evaporation will end up entangled with the interior state before total evaporation, even when the black hole and everything confined within it has vanished – i.e. the initial pure state evolved into a final mixed state \cite{Unruh:2017uaw}. Moreover, once evaporation is complete the spacetime will return to its vacuum state, which has traditionally been regarded as unique and incapable of storing nontrivial information \cite{Strominger:2017zoo}. It is this violation of unitarity that has placed the information paradox among the most enigmatic open problems in modern physics.\footnote{It is important to distinguish between this class of unitarity violation and the breakdown of unitarity in more general settings. In fact, one could observe a non-unitary evolution by considering a quantum system on a Cauchy surface at an early stage and a hyperbolic surface at a late stage \cite{Unruh:2017uaw}. Since the late-time surface is non-Cauchy, information is lost over the course of the system’s evolution. The essential difference being that, in the latter context, information loss arises out of ignorance and not as a consequence of a fixed physical circumstance - i.e. we chose to consider the system’s final state on a hyperbolic surface, had we chosen a Cauchy surface instead (which we were free to do) we would have observed a unitary evolution. The same choice cannot be made in the context of black holes, since the lost information has irrevocably vanished.} \\

One is left with two options at this point which can be succinctly summarised in terms of a collection of ideas that have become known as the "central dogma" of black hole physics. According to the dogma, an external observer can describe a black hole as a quantum system whose degrees of freedom are proportional to its area and whose evolution is unitary \cite{Almheiri:2020cfm}. In the absence of a clear mechanism for the flow of information out of a black hole, one must accept that either the central dogma is false and nonunitary evolutions are a feature of quantum gravity or we are missing something substantial in our survey of black hole physics thus far.\footnote{Taking nonunitary evolutions to be a feature of a quantum theory of gravity is a somewhat radical, albeit not inconceivable, proposal since it leaves us in a universe fundamentally bereft of deterministic laws and causal order. In fact, unitarity implies a system is both deterministic and time-reversible. If quantum gravity is nonunitary then one cannot have complete information about a system at some moment and predict its future, nor its past.} Recent developments by Perry, Hawking and Strominger have taken aim towards the latter. Their work, which examined soft electric and gravitational charges, found that the infinite set of conservation laws obeyed by a soft charge species necessarily implants additional (soft) hair on black holes {Hawking:2016msc}, {Hawking:2016sgy}. In fact, soft hairs can be envisioned as pixels on a holographic plate that lives at the future boundary of a black hole horizon. Any particle that crosses the horizon will excite a pixel on the holographic plate. Hawking's original argument for information loss in black holes is therefore invalid, and a generic black hole can carry infinitely more observables than previously thought \cite{Hawking:2016msc}. The fact that an infinity of soft charges live in and characterise the vacuum states of any gauge or gravitational theory also reveals another flaw in prior investigations of the paradox: for an otherwise empty spacetime containing a black hole, the vacuum that the spacetime settles down to after black hole evaporation is not generally the same as the vacuum state prior to turning on evaporation. A considerable amount of information that the black hole carried is therefore not lost completely but is rather, in some way, encoded into the vacuum structures of the theory – in general, into the electromagnetic, chromodynamic, and gravitational vacua.  \\

Generalising this setup to account for both abelian and nonabelian soft hairs is straightforward. Let $|B \rangle$ denote the quantum state of a black hole defined on a horizon $\mathcal{H}$. In any gravitational scattering process, such as the formation and evaporation of a black hole, $|B \rangle$ can be regarded as the incoming state. If the state contains only neutral matter then the current on $\mathcal{H}$ vanishes and, from \eqref{eq: Soft Charges as Operators}, we have

\begin{gather}\label{eq: Soft Charges on Black Holes as Operators}
\mathcal{Q}_{\epsilon}|B \rangle = \left( \int_{\partial \Sigma} \Lambda \star F \right)|B \rangle \neq  0 \\\
\mathcal{Q}_{m}|B \rangle = \left( \int_{\partial \Sigma} \Lambda F \right)|B \rangle \neq  0 \label{eq: Soft Charges on Black Holes as Operators II}
\end{gather}\  \enlargethispage{\baselineskip} 

where $\mathcal{Q}_{\epsilon}$ and $\mathcal{Q}_{m}$ may refer to the soft electric and magnetic charges of Maxwell's theory or to the soft chromoelectric and chromomagnetic charges of Yang-Mills theory respectively.\footnote{To simplify the form of \eqref{eq: Soft Charges on Black Holes as Operators} and \eqref{eq: Soft Charges on Black Holes as Operators II} we have taken the trace operator to be implicit for the Yang-Mills case.} The state $\mathcal{Q}|B \rangle$ is the black hole state $|B \rangle$ together with a soft (gluon) photon with polarisation vector proportional to $\mathcal{D}\Lambda$. If we label this state as $|B' \rangle$, then we have that $|B' \rangle$ and $|B \rangle$ are the same state up to a soft gauge particle and are therefore energetically-degenerate. \\

Assuming unitarity, after a finite time $|B \rangle$ would evaporate into some pure state $|X\rangle$ supported on $\mathcal{I}^{+}$ whereas $|B' \rangle$ would evaporate into a pure state $|X'\rangle$ supported on $\mathcal{I}^{+}$ \cite{Hawking:2016msc}.\footnote{We emphasize that there is no known algorithm that would enable us to determine precisely which state the system will end up in after evaporation, presenting a major hurdle towards a complete resolution of the paradox \cite{Hawking:2016msc}.} If we now divide the spacetime containing the black hole with a spacelike hypersurface that separates the complete horizon formation process from the period where evaporation begins, then we can take $u_{0}$ to be the (retarded) time that the hypersurface intersects $\mathcal{I}^{+}$ and $v_{0}$ to be the (advanced) time that the hypersurface intersects $\mathcal{H}$. We are therefore left with a region of future null infinity prior to evaporation, which we label $\mathcal{I}^{+}_{<}$, and a region associated to the complete evaporation process labelled by $\mathcal{I}^{+}_{>}$. Similarly, we denote the portion of the horizon before evaporation by $\mathcal{H}_{<}$. We impose the additional simplification that the portion of the spacelike slice between the horizon and future null infinity is empty, so that  $\mathcal{I}^{+}_{<} \cup \mathcal{H}_{<}$ contains the complete Cauchy data of the system (i.e. $\mathcal{I}^{+}_{<} \cup \mathcal{H}_{<}$ is a Cauchy surface). \\

Since we've assumed that no currents are present, there are no charges present in the distant (timelike) future $i^{+}$. We can thus write \cite{Hawking:2016msc}

\begin{equation}\label{eq: Soft Charge Conserv}
\mathcal{Q}\, =\, \mathcal{Q}^{\mathcal{I}^{+}} = \, \mathcal{Q}^{\mathcal{I}^{+}_{<}} + \, \mathcal{Q}^{\mathcal{I}^{+}_{>}}
\end{equation}\ 

which is nothing but an expression of soft charge conservation in terms of the regions of null infinity ``before" and ``during" the evaporation process. Since $\mathcal{I}^{+}_{<} \cup \mathcal{H}_{<}$ is Cauchy, we may also write \cite{Hawking:2016msc}

\begin{equation}\label{eq: Soft Charge Conserv w Horizon}
\mathcal{Q} = \mathcal{Q}^{\mathcal{I}^{+}_{<}} + \mathcal{Q}^{\mathcal{H}_{<}}
\end{equation}\

which captures all of the soft charge between $\mathcal{I}^{+}_{<}$ and $\mathcal{H}_{<}$. By direct comparison of \eqref{eq: Soft Charge Conserv} and \eqref{eq: Soft Charge Conserv w Horizon} it follows immediately that the soft charge on $\mathcal{H}_{<}$ is subsequently supported on $\mathcal{I}^{+}_{>}$. The incoming states $|B \rangle$ and $|B' \rangle$ defined on $\mathcal{H}_{<}$ and related by a soft charge therefore evolve into outgoing states $|X \rangle$ and $|X' \rangle$ defined on $\mathcal{I}^{+}_{>}$ that obey \cite{Strominger:2017zoo}

\begin{equation}\label{eq: Soft Charge Conserv w Horizon II}
|X' \rangle = \mathcal{Q}^{\mathcal{I}^{+}_{>}}|X\rangle
\end{equation}\

Since there is an infinity of possible choices of soft charge that we could have used in each of the electromagnetic and Yang-Mills cases, corresponding to a different choice of $\Lambda$, we see that there are an infinite number of exact, deterministic relations between the outgoing state arising from the black hole state $|B \rangle$ and the outgoing state arising out of the soft charged, or large-gauge transformed, state $|B' \rangle$. In short, the process of black hole formation and evaporation is tightly constrained by the existence and conservation of soft charges \cite{Hawking:2016msc}.\\

\section{Conclusion}

We have found that an infinity of soft magnetic charges in Maxwell's theory arise independently of the presence of magnetic sources. This construction was then generalised to derive an expression for the two infinite sets of soft charges existing in Yang-Mills theory. The abelian and nonabelian soft charges were subsequently related to physically significant gauge symmetries of the electromagnetic and Yang-Mills fields respectively. Finally, these results were applied to the black hole information paradox where we argued that the deep infrared anatomy of a black hole is comprised of not only an infinite amount of soft gravitational hair but also of an infinite number of abelian and nonabelian soft hairs. While this has not yet provided a solution to the paradox, it has yielded important insights into the nature of information loss in black hole physics. There are still several unknowns that remain, chiefly among which is an investigation into whether the information describing in-falling matter fields can be encoded in the soft hairs of a black hole. We suspect that a potential way forward may lie within a formulation of black hole soft implants in string theory since the duality between string theory and gauge theory implies that all fields can be regarded as gauge fields \cite{Ooguri:1999tx}. From this point of view, it is conceivable that the information associated with matter fields may be stored in ``matter" soft hairs. It is desirable, however, to obtain an algorithm for information flow via soft hair that is independent of any purported theory of quantum gravity, which we leave for future research.

\section*{Acknowledgements}
I would like to thank my supervisor and mentor, Malcolm John Perry, for his indispensable advice and for the knowledge and expertise that he afforded the development of this paper.

\end{document}